\documentclass[prl,twocolumn,showpacs,amsmath,amssymb,amsfont]{revtex4}

\usepackage{graphicx}

\begin{document}

\title{Quantum phase transition in superconducting Au$_{0.7}$In$_{0.3}$
films of very low normal-state sheet resistance}

\author{M.~M. Rosario}

\author{H. Wang}

\author{Yu. Zadorozhny}

\author{Y. Liu}

\affiliation{Department of Physics, The Pennsylvania State University,
University Park, PA 16802}

\date{\today}

\begin{abstract}

We report the observation of a quantum phase transition (QPT), tuned by a
parallel magnetic field, between a superconducting and metallic state in
Au$_{0.7}$In$_{0.3}$ films of very low normal-state sheet resistance ($<
90~\Omega $). These films can be modeled as a random array of
superconductor-normal metal-superconductor (SNS) junctions. Electrical
transport and tunneling measurements suggest that, in the metallic state,
the film consists of superconducting In-rich grains not linked by
Josephson coupling. Whether phase fluctuation, which is not expected to be
strong in such an SNS junction system according to the phase--number
uncertainty relation, or a different physical process drives the observed
QPT is discussed.

\end{abstract}


\maketitle

% Intro

As an example of a quantum phase transition (QPT), the
superconductor-insulator transition (SIT) in two dimensions (2D) has been
an important subject of study in contemporary condensed matter
physics~\cite{markovicreview}. Consideration of the 2D SIT observed in
granular films~\cite{jaeger86,dynes86} and
superconductor-insulator-superconductor (SIS) Josephson junction
arrays~\cite{chenandvanderzant} usually starts from quantum phase
fluctuations. The SIT observed in homogenous
films~\cite{haviland89,hebard90,valles92} has been analyzed in a dirty
Bose-Hubbard model~\cite{fisher90} that builds on phase considerations as
well. The physical origin of the phase fluctuation is captured by an
uncertainty relation between the superconducting phase $\phi $ and the
number of carriers $N$ with the form $\Delta \phi \Delta N \approx 1$. For
SIS junction arrays or granular films, the large charging energy of the
superconducting islands makes the transfer of Cooper pairs between
neighboring islands difficult, suppressing the fluctuation in $N$. As a
result, the fluctuation in the phase is enhanced, leading ultimately to an
SIT. The introduction of shunt resistance or dissipation tends to suppress
the phase fluctuation. In recent experiments, it was found that
dissipation can restore the global phase coherence for a system sitting on
the insulating side of the SIT~\cite{takahide,clarke}, as
anticipated~\cite{chakravartyandkivelson}. For amorphous films, the
localization of electrons by disorder can reduce $\Delta N$, and
consequently lead to enhanced phase fluctuation and an SIT. With no
exception, the normal-state sheet resistance $R_{\Box}^{\rm N}$ of the
films around the SIT is large, typically around $h/4e^2 = 6.45~{\rm
k}\Omega$~\cite{markovicreview}.

Recently, the possible existence of a metallic state in 2D SIT systems has
received renewed theoretical
attention~\cite{doniachandphillips,kapitulnik01}. The apparent existence
of such a state occuring between the insulating and the superconducting
phase, marked by a flat resistance tail at the lowest temperatures, was
seen in ultrathin granular films of Ga, Pb, and In measured down to
0.6~K~\cite{jaeger86}, in thin Al films down to 0.3~K~\cite{wu94}, and in
SIS Josephson junction arrays down to 10~mK~\cite{chenandvanderzant}.
Based on experiments on amorphous MoGe films in perpendicular magnetic
field~\cite{mason99}, it was proposed that dissipation may open a
parameter space for a metallic state to occur near a 2D QPT in
general~\cite{kapitulnik01}. In all these previous studies, the metallic
state was found in systems with large $R_{\Box}^{\rm N}$ ($> 1~{\rm
k}\Omega$). Here we report electrical transport and tunneling measurements
in Au$_{0.7}$In$_{0.3}$ thin films, showing the existence of a QPT and a
metallic state in films with very low $R_{\Box}^{\rm N}$ ($< 90~\Omega $).

% Experimental

Au$_{0.7}$In$_{0.3}$ films were prepared by sequential thermal evaporation
of alternating Au and In layers, with the layer thicknesses determined by
the desired atomic ratio of Au to In. The maximum solid solubility of In
in Au is around 10\% at room temperature. The interdiffusion of Au and In
in a Au$_{0.7}$In$_{0.3}$ film results in a 2D system consisting of
In-rich grains, with a maximum local $T_{\rm c}$ of 0.6--0.8K, embedded in
a Au$_{0.9}$In$_{0.1}$ matrix, with $T_{\rm c}$ around
77~mK~\cite{zador02}. Since the atomic composition and the size of the
In-rich grains can vary randomly, strong spatial variation in the
amplitude of the superconducting order parameter (the superconducting gap)
is expected. It has been shown previously that this system can be modeled
as a random array of superconductor-normal metal-superconductor (SNS)
Josephson junctions~\cite{zador02}.

Planar tunnel junctions of Au$_{0.7}$In$_{0.3}$/MgO$_x$/Mg were made in a
standard cross geometry, with a junction size of $0.2 \times 0.3~{\rm
mm}^2$. The Mg bottom layer was thermally evaporated at ambient
temperature, with subsequent growth of a native Mg oxide layer encouraged
by the use of glow discharge in an O$_2$ environment. The deposition of
the Au$_{0.7}$In$_{0.3}$ top layer was carried out with the substrate held
at liquid nitrogen temperatures ($\approx 77$~K) to help preserve the
insulating barrier.  The results presented here correspond to a junction
with a normal state resistance of $115~\Omega$. Measurements on another
junction of comparable resistance and two junctions of higher resistance
($\sim 10^3 \Omega$) yielded qualitatively similar results.

Electrical transport measurements were carried out in a dilution
refrigerator equipped with a superconducting magnet. The base temperature
was $<20$~mK. All electrical leads entering the cryostat were filtered
with the attenuation of 10~dB at 10~MHz and 50~dB at 300~MHz. Resistances
and current-voltage ($I-V$) characteristics were measured with a d.c.
current source and a nanovoltmeter. Tunneling conductances $G_{\rm j}(V)$
were determined by taking the derivative of $I-V$ curves numerically. The
magnetic field was applied parallel to the film plane (estimated to be
aligned within about $1^{\rm o}$) and perpendicular to the tunneling
direction.

% Results - metallic state, transport

Figure~1 shows the temperature dependence of the sheet resistance
$R_{\Box}(T)$ in parallel magnetic field $H_{\parallel}$ for several
Au$_{0.7}$In$_{0.3}$ films. The normal-state sheet resistance of these
films, ranging from 10--90~$\Omega $, are very low compared with $h/4e^2$
where the 2D SIT typically occurs. The superconducting transition
temperature $T_{\rm c}$ was found to decrease with increasing
$H_{\parallel}$. For relatively small $H_{\parallel}$, a fully
superconducting state was obtained. However, despite the presence of a
substantial resistance drop slightly below the zero-field $T_{\rm c}$,
zero resistance was not reached down to $T = 20$~mK as $H_{\parallel}$
surpassed a critical value $H_{\parallel }^{\rm c}$. Instead, a flat
resistance tail was found, spanning nearly a decade in temperature. The
limiting ($T \to 0$) resistance increased exponentially with
$H_{\parallel}$ (Fig.~2).

Figure~3 shows $I-V$ characteristics for the 10~nm thick film (shown in
Fig.~1b). Data presented in Fig.~1b were obtained at currents of 1$\mu $A
or 100nA, with no qualitative differences in $R(T)$ behavior. The $I-V$
characteristic evolved from nonlinear to linear (ohmic) behavior with
increasing $H_{\parallel}$ at the lowest temperatures (Fig. 3b). According
to the Kosterlitz-Thouless (KT) theory~\cite{ktandhalperin}, the finite
temperature superconducting transition in 2D is associated with the
thermal unbinding of vortex-antivortex pairs, leading to a $I-V$
characteristic of $V\sim I^3$ at $T=T_{\rm KT}$. In these
Au$_{0.7}$In$_{0.3}$ films, the exponent was found to be less than 3 down
to the lowest temperature ($V\sim I^{2.5.}$), perhaps due to a vanishing
$T_{\rm KT} < 25$~mK. Despite this, the non-linear $I-V$ characteristic at
the lowest temperatures indicates that vortices and antivortices were
present at $H_{\parallel}=0$. Linear $I-V$ characteristics were found even
at 25~mK at $H_{\parallel} = 0.20$T, above which the resistance tail
emerged, suggesting the absence of vortex-antivortex unbinding in the
metallic state. However, whether vortices and antivortices were absent, or
were present but fully unbound, was not resolved.

% Results - tunneling

The interesting question is whether In-rich grains remain superconducting
in the metallic state. Tunneling measurements were carried out to address
this. Single particle tunneling spectra, obtained in
Au$_{0.7}$In$_{0.3}$/MgO$_x$/Mg junctions at various temperatures and
applied parallel fields, are shown in Fig.~4. The tunneling spectra in
zero field indicates that an energy gap opened below $T = 0.28$~K. While a
coherence peak is present in the tunneling spectra, the peak is smaller
and the zero bias conductance $G_{\rm j}(V=0)$ is larger than expected
from BCS theory. A 35\% suppression of the normal state DOS was observed
at 20~mK for junction shown in Fig.~4a and 25--80\% for others at the zero
bias, resulting from either a substantial population of quasiparticles in
Au$_{0.7}$In$_{0.3}$ or junction leakage. The superconducting gap $\Delta
_0$ is estimated by the peak position to be 0.1~meV at 25~mK, and is
smaller if one uses $G_{\rm j}(V = \Delta /e) = G_{\rm j}^{\rm N}$. With
$T_{\rm c}^{\rm onset}=0.82$~K, this leads to $\Delta _0/{\rm k_B}T_{\rm
c} = 1.41$, slightly smaller than the BCS result, $\Delta /{\rm k_B}T_{\rm
c} =1.76$.

The evolution of the tunneling spectra with $H_{\parallel}$ is shown in
Fig.~4b. With increasing field, more states were found within the gap
until it closed at $H_{\parallel }\approx 0.55$~T. With increasing
$H{\parallel }$, the coherence peak decreased in height and broadened in
width. The zero-bias conductance, which could be related to the total area
of the normal region of the films, increased linearly with $H_{\parallel}$
(data not shown).

A natural picture concerning the observed superconducting-metallic state
transition in Au$_{0.7}$In$_{0.3}$ films emerges from these measurements.
With the application of $H_{\parallel}$, the superconductivity in the
In-rich grains is gradually reduced. Eventually, a sufficient number of
grains become normal so that a percolating path of Josephson coupled
superconducting grains can no longer form, leading to the disappearance of
global superconductivity. The average separation between superconducting
islands at $H_{\parallel}^{\rm c}$ can be estimated. Finite Josephson
coupling of an SNS junction is expected if the length of the N-layer is
shorter than a few times of the normal coherence length $\xi _{\rm N}$. In
the dirty limit, $\xi _{\rm N} = (\hbar D/2\pi {\rm k_B}T)^{1/2}$ where $D
= v_{\rm F}\tau $ is the diffusion constant, $v_{\rm F}$ is the Fermi
velocity, and $\tau $ is the relaxation time in the Boltzmann formula for
resistivity, $\rho = m/ne^2\tau $. For the film shown in Fig.~1b, for
example, we estimate $\xi _{\rm N}$ for the normal metal matrix
(Au$_{0.9}$In$_{0.1}$) to be $\approx 0.2~\mu $m at 20~mK. This suggests
that the average separation between surviving superconducting islands is
in the micron range, comparable to the average size of the largest In-rich
grains.

Recently, Larkin and co-workers have proposed~\cite{feigelman01} a theory
for an SNS array of superconducting islands (of radius $d$ and spaced $b$
apart, such that $b \gg d$) proximity coupled to one another via a 2D
normal film of dimensionless conductance $g = \sigma /(e^2/\hbar )$. In
their model, quantum phase fluctuations induced by disorder and Coulomb
repulsion are responsible for the suppression of superconductivity. A QPT
from a superconducting to a normal-metal state was shown to occur at $g <
g_{\rm c} \approx [(1/\pi){\rm ln}(b/d)]^2$. The corresponding critical
sheet resistance $R_{\Box \rm c}$ can be substantially smaller than $R_Q$.
For the films shown in Fig.~1, we estimate $R_{\Box \rm c} \approx
R_{\Box}^N$. Then $R_{\Box}^N = 9.96\Omega $ would yield $g_{\rm c}= 404$.
An unreasonably large distance between grains, given by $b = d{\rm
exp}(63)$, would be obtained at the QPT, larger than the value inferred
from the experiment, as described above. This appears to suggest that some
important ingredients are missing from this model in its present form.

An alternative model has been proposed in which the effects of
fluctuations in the amplitude of the superconducting order parameter,
primarily as a function of time, are taken into account~\cite{spivak01}.
Presumably, the amplitude fluctuation will be present when the radius of
the superconducting island $d$ is smaller than the zero-temperature
superconducting coherence length $\xi _0$, such that $d < \xi _0$. In the
dirty limit, $\xi _0 = (\hbar D/\Delta )^{1/2}$~\cite{feigelman01}. This
yields $\xi _0 \approx 0.1\mu $m for the present study. The size of many
superconducting grains in Au$_{0.7}$In$_{0.3}$ films is expected to be
smaller than this~\cite{zador02}, making substantial amplitude fluctuation
plausible. The critical concentration of grains obtained in this model is
substantially larger than that obtained in Ref.~\cite{feigelman01},
leading to a more reasonable value of critical conductance, qualitatively
consistent with our experimental observation. Unfortunately, quantitative
predictions are lacking, making a quantitative comparison between the
experiment and the theory impossible.

It was previously emphasized that amplitude fluctuations played a role in
the 2D SIT in ultrathin amorphous films~\cite{valles92,hsu95}. The
vanishing of the gap~\cite{valles92}, the reduced superconducting
condensation energy~\cite{hsu95}, as well as the broadening of the
tunneling spectrum~\cite{hsu95} near the SIT were cited as the evidence
for amplitude fluctuations. However, while the vanishing of the gap and
condensation energy create favorable conditions for the
amplitude of the order parameter to fluctuate, direct experimental
evidence for this near a $T=0$ SIT is yet to be found. Given this, it
might be helpful to consider possible driving forces for the amplitude
fluctuation.

The simplest consideration suggests that a deparing process is needed to
induce amplitude fluctuation. The residual repulsive electron-electron
interaction appears to be an obvious driving force for amplitude
fluctuation. Theoretically, a magnetic field applied parallel to a
homogeneously disordered 2D superconducting system was shown specifically
to lead to strong amplitude fluctuations due to Zeeman splitting and
 strong spin-orbit coupling~\cite{zhou98}. In this context, physical
phenomena which may result from the presence of negative superfluid
density ({\it i.e.}, $\pi $ junctions with negative Josephson coupling),
an extreme case of the amplitude fluctuation, have been observed in
Au$_{0.7}$In$_{0.3}$ cylindrical films~\cite{zador01}.

The observation of a flat resistance tail in films with such low $R_{\Box
}^{\rm N}$ is striking. Is it possible that the tail originates from
electrons being at a temperature higher than the lattice temperature
because of insufficient cooling? It has been emphasized that a flat
resistance tail has never been observed in granular or amorphous films
prepared in some laboratories~\cite{valles92,hsu95,dynes02}. In the
present work, noise from outside the system were eliminated by RF filters.
However, microwave noise originating from room-temperature parts of the
measuring leads inside the cryostat may still affect the film resistance.
Effective elimination of these noises requires filtering at cryogenic
temperatures, for which the current system is not equipped. On the other
hand, a similar phenomenon in Josephson junction arrays was observed in
studies carried out with such filters~\cite{chenandvanderzant}. In
addition, the films in the present study were of very low $R_{\Box }^{\rm
N}$, which should make heating due to the electromagnetic environment less
significant.

Nevertheless, it is desirable to show directly that electrons still cool
in the temperature range in which the flat resistance tail was seen.
Figure~5a shows the tunneling spectra at 20 and 50mK in zero field,
indicating that the electronic system still cooled down to the lowest
temperatures. Above $H_{\parallel}^{\rm c}$, the spectra became less
distinguishable. However, this could be due to depairing effects. For a
cylindrical film above $H_{\parallel }^{\rm c}$, the resistance shows a
negative d$R$/d$T$ (Fig. 5c), indicating that the electrons were cooling
in the metallic state for this sample. However, $R(T)$ at 5~T for the
7.5nm thick film showed a change of slope or flattening-off at low
temperatures (Fig.~5d). Whether this was because superconducting
fluctuations were still present or electrons were not cooling was not
resolved. Direct measurements of the electronic temperature are needed to
clarify this issue.

In conclusion, we would like to thank S. Girvin, A.M. Goldman, S.
Kivelson, P. Phillips, J. Rowell , J. Valles, Jr., and F. Zhou for useful
discussions on various aspects of this work. We acknowledge support from
NSF through Grant DMR 0202534.

\begin{figure}

\caption{\label{fig1} Sheet resistance as a function of temperature,
$R_{\Box}(T)$, of superconducting Au$_{0.7}$In$_{0.3}$ films of several
thicknesses, $t$, as indicated. Planar films with (a)~$R_{\Box }^{\rm
N}=89.2\Omega $, (b)~$R_{\Box }^{\rm N}=55.7\Omega $, (c)~$R_{\Box }^{\rm
N}=9.90\Omega $, and (d)~a cylindrical film with diameter $d=550$~nm and
$R_{\Box }^{\rm N}=9.96\Omega $ are shown.}

\end{figure}

\begin{figure}

\caption{\label{fig2} Semi-log plot of the limiting ($T \to 0$) resistance
normalized to the zero field normal state resistance, $R/R^{\rm N}$, as a
function of applied parallel magnetic field normalized to the field
required to suppress the resistance drop in $R(T)$,
$H_{\parallel}/H_{\parallel }^{\rm N}$, for the films shown in Fig.~1. The
film given in Fig.~1c, not included in the plot as data above 0.2T is not
available, showed similar behavior.}

\end{figure}

\begin{figure}

\caption{\label{fig3} (a)~$I-V$ characteristic of the 10~nm thick
Au$_{0.7}$In$_{0.3}$ film at $H_{\parallel} =0$. Curves are given for
$T=20$~mK, 90~mK, 0.1~K, 0.125~K, 0.175~K, and 0.20~K. The dashed line
indicates linear (ohmic) behavior. The low current region of the 20~mK
curve follows $V\sim I^{\alpha }$ where $\alpha \approx 2.5$. (b)~$I-V$
characteristic at finite $H_{\parallel}$. Curves were taken at the fields
indicated and at $T=25$~mK, with the exception of the 0.1~T curve which
was taken at 70~mK.}

\end{figure}

\begin{figure}

\caption{\label{fig4} Tunneling spectra of the 10~nm thick
Au$_{0.7}$In$_{0.3}$ film. (a)~Tunneling conductance as a function of
voltage, $G_{\rm j}(V)$, at several temperatures at $H_{\parallel }=0$.
$T>25$~mK curves are shifted up from the 25~mK curve for clarity.
(b)~$G_{\rm j}(V)$ curves at finite $H_{\parallel}$ and $T=25$mK.
$H_{\parallel}>0$ curves are shifted up from the $H_{\parallel }=0$ curve.
The high energy features [{\it e.g.}, at 0.2~mV for 0.28~K in (a)] appears
to be related to a junction-dependent current redistribution process and
is not intrinsic to Au$_{0.7}$In$_{0.3}$.}

\end{figure}

\begin{figure}

\caption{\label{fig5} Tunneling spectra of the 10~nm thick
Au$_{0.7}$In$_{0.3}$ film for $T=25$ and 50~mK at (a)~$H_{\parallel}=0$
and (b)~0.23T. (c)~$R_{\Box }(T)$ of a cylindrical film with $t=30$nm and
$d=470$nm in $H_{\parallel } = 0.26$T and 0.6T. (d)~A close-up of the
$R(T)$ of the 7.5nm film shown in Fig.~1a.}

\end{figure}

\end{document}